\documentclass[12pt,preprint]{aastex}
\newcounter{address}
\newcommand{\documentname}{\textsl{Article}}
\begin{document}

\title{What bandwidth do I need for my image?}
\author{
  Adrian~M.~Price-Whelan\altaffilmark{\ref{CCPP}},
  David~W.~Hogg\altaffilmark{\ref{CCPP},\ref{email}}
}

\altaffiltext{\theaddress}{\stepcounter{address}\label{CCPP} Center
for Cosmology and Particle Physics, Department of Physics, New York
University, 4 Washington Place, New York, NY 10003}
\altaffiltext{\theaddress}{\stepcounter{address}\label{email} To whom
correspondence should be addressed: \texttt{david.hogg@nyu.edu}}

\begin{abstract}
Computer representations of real numbers are necessarily discrete,
with some finite resolution, discreteness, quantization, or minimum
representable difference.  We perform astrometric and photometric
measurements on stars and co-add multiple observations of faint
sources to demonstrate that essentially all of the scientific
information in an optical astronomical image can be preserved or
transmitted when the minimum representable difference is a factor of
two finer than the root-variance of the per-pixel noise. Adopting a
representation this coarse reduces bandwidth for data acquisition,
transmission, or storage, or permits better use of the system
dynamic range, without sacrificing any information for down-stream
data analysis, including information on sources fainter than the
minimum representable difference itself.
\end{abstract}


\section{Introduction}
Computers operate on bits and collections of bits; the numbers stored
by a computer are necessarily discrete; finite in both range and
resolution.  Computer-mediated measurements or quantitative
observations of the world are therefore only approximately
real-valued.  This means that choices must be made, in the design of a
computer instrument or a computational representation of data, about
the range and resolution of represented numbers.

In astronomy this limitation is keenly felt at the present day in
optical imaging systems, where the analog-to-digital conversion of CCD
or equivalent detector read-out happens in real time and is severely
limited in bandwidth; often there are only eight bits per readout
pixel.  This is even more constrained in space missions, where it is
not just the bandwidth of real-time electronics but the bandwidth of
telemetry of data from space to ground that is limited.  If the
``gain'' of the system is set too far in one direction, too much of
the dynamic range is spent on noise, and bright sources saturate the
representation too frequently.  If the gain is set too far in the
other direction, information is lost about faint sources.

Fortunately, the information content of any astronomical image is
limited \emph{naturally} by the fact that the image contains
\emph{noise}.  That is, tiny differences between pixel
values---differences much smaller than the amplitude of any additive
noise---do not carry very much astronomical information.  For this
reason, the discreteness of computer representations of pixel values
do not have to limit the scientific information content in a
computer-recorded image.  All that is required is that the noise in
the image be \emph{resolved} by the representation.  What this means,
quantitatively, for the design of imaging systems is the subject of
this \documentname; we are asking this question: ``What bandwidth is
required to deliver the scientific information content of a
computer-recorded image?''

This question has been asked before, using information theory, in the
context of telemetry \citep{Gaztanaga} or image compression
\citep{Watson}, treating the pixels (or linear combinations of them)
as independent.  Here we ask this question, in some sense,
\emph{experimentally}, and for the properties of imaging on which
optical astronomy depends, where groups of contiguous pixels are used
in concert to detect and centroid faint sources.  We perform
experiments with artificial data, varying the bandwidth of the
representation---the size of the smallest representable difference
$\Delta$ in pixel values---and measuring properties of scientific
interest in the image.  We go beyond previous experiments of this kind
(\citealt{WhiteGreenfield}, and \citealt{PenceInPress}) by
measuring the centroids and brightnesses of compact sources, and
sources fainter than the detection limit.  The higher the bandwidth,
the better these measurements become, in precision and in accuracy.
We find, in agreement with previous experiments and
information-theory-based results, that the smallest representable
difference $\Delta$ should be on the order of the root-variance
$\sigma$ of the noise in the image.  More specifically, we find that
\emph{the minimum representable difference should be about half the
  per-pixel noise sigma} or that about two bits should span the FWHM
of the noise distribution if the computer representation is to deliver
the information content of the image.

Of course, tiny mean differences in pixel values, even differences
much smaller than the noise amplitude, \emph{do} contain
\emph{extremely valuable} information, as is clear when many short
exposures (for example) of one patch of the sky are co-added or
analyzed simultaneously.  ``Blank'' or noise-dominated parts of the
individual images become signal-dominated in the co-added image.  In
what follows, we explicitly include this ``below-the-noise''
information as part of the information content of the image.  Perhaps
surprisingly, \emph{all} of the information can be preserved, even
about sources fainter than the discreteness of the computer
representation, provided that the discreteness is finer than the
amplitude of the noise.  This result has important implications for
image compression, but our main interest here is in the design and
configuration of systems that efficiently take or store raw data,
using as much of the necessarily limited dynamic range on signal as
possible.

Our results have some relationship to the study of \emph{stochastic
  resonance}, where it has been shown that signals of low dynamic
range can be better detected in the presence of noise than in the
absence of noise (see \citealt{stochres} for a review).  These studies
show that if a signal is below the minimum representable difference
$\Delta$, it is visible in the data only when the digitization of the
signal is noisy.  A crude summary of this literature is that the
optimal noise amplitude is comparable to the minimum representable
difference $\Delta$.  We turn the stochastic resonance problem on its
head: The counterintuitive result (in the stochastic resonance
context) that weak signals become detectable only when the
digitization is noisy becomes the relatively obvious result (in our
context) that so long as the minimum representable difference $\Delta$
is comparable to or smaller than the noise, signals are transmitted at
the maximum fidelity possible in the data set.

What we call here the ``minimum representable difference'' has also
been called by other authors the ``discretization'' \citep{Gaztanaga}
or ``quantization'' \citep{Watson, WhiteGreenfield, Pence}.

\section{Method}

The artificial images we use for the experiments that follow are all
made with the same basic parameters and processes.  The images are
square $16\times 16$ pixel images, to which we have added Gaussian
noise to simulate sky (plus read) noise.  A random number generator
chooses a mean sky level $\nu$ for this Gaussian noise within the
range 0 to 100 but the variance $\sigma^2$ is fixed for all
experiments at $\sigma^2 = 1$.  For most experiments we also add a
randomly placed ``star'' with a Gaussian point-spread function at a
location $(x_0,y_0)$ within a few pixel radius of the center of the
image. The intensity of the star is given by a circular 2-dimensional
Gaussian function. The FWHM of the star point-spread function is set
to 2.35~pix for convenience, and the total flux of the star---total
counts above background after integration over the array---is a
variable.  In the experiments to follow we set this total flux $S$ to
2.0, 64.0, and 2048.0.  Given the FWHM setting of 2.35~pix and the sky
noise setting of $\sigma^2=1$, the peak intensities corresponding to
these fluxes are $0.32\,\sigma$, $10\,\sigma$, and $320\,\sigma$.

When we add the star, we do not add any Poisson or star-induced noise contribution to the images.  That is, the images are ``sky-dominated'' in the sense that the per-pixel noise is the same in the center of the star as it is far from the star.  In the context of setting the minimum representable difference, this choice is conservative, but it is also slightly unrealistic.

\begin{figure}
\includegraphics[width=\textwidth]{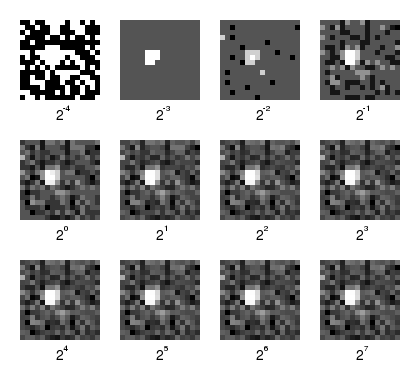}
\caption{Starting from top left and moving to bottom right we show
  $16\times 16$ images of increasing bit depth. The original images are
  identical but snapped to integer as described in the text. The
  images are labeled by the ratio $[\sigma/\Delta]$ of noise
  root-variance $\sigma$ to the minimum representable difference
  $\Delta$. At ratios $[\sigma/\Delta] > 2^{0}$, the
  images become virtually indistinguishable from the high bandwidth
  images.\label{fig:twelvepanel}}
\end{figure}

The method for setting the minimum representable difference $\Delta$
for the artificial images is used extensively in the experiments to
follow. We define an array of factors $\Delta$ ranging from
$2^{4}\,\sigma$ to $2^{-8}\,\sigma$, which we use to scale the
data. We divide the image data by each value of $\Delta$, round all
pixels to their nearest-integer values, and then multiply back in by
$\Delta$.  For convenience, we will call this ``scale, snap to
integer, un-scale'' procedure
``SNIP''. \figurename~\ref{fig:twelvepanel} shows identical images
SNIPped at different multiplicative factors; each panel shows the same
original image data but with a different minimum representable
difference $\Delta$.

\begin{figure}
\includegraphics[width=\textwidth]{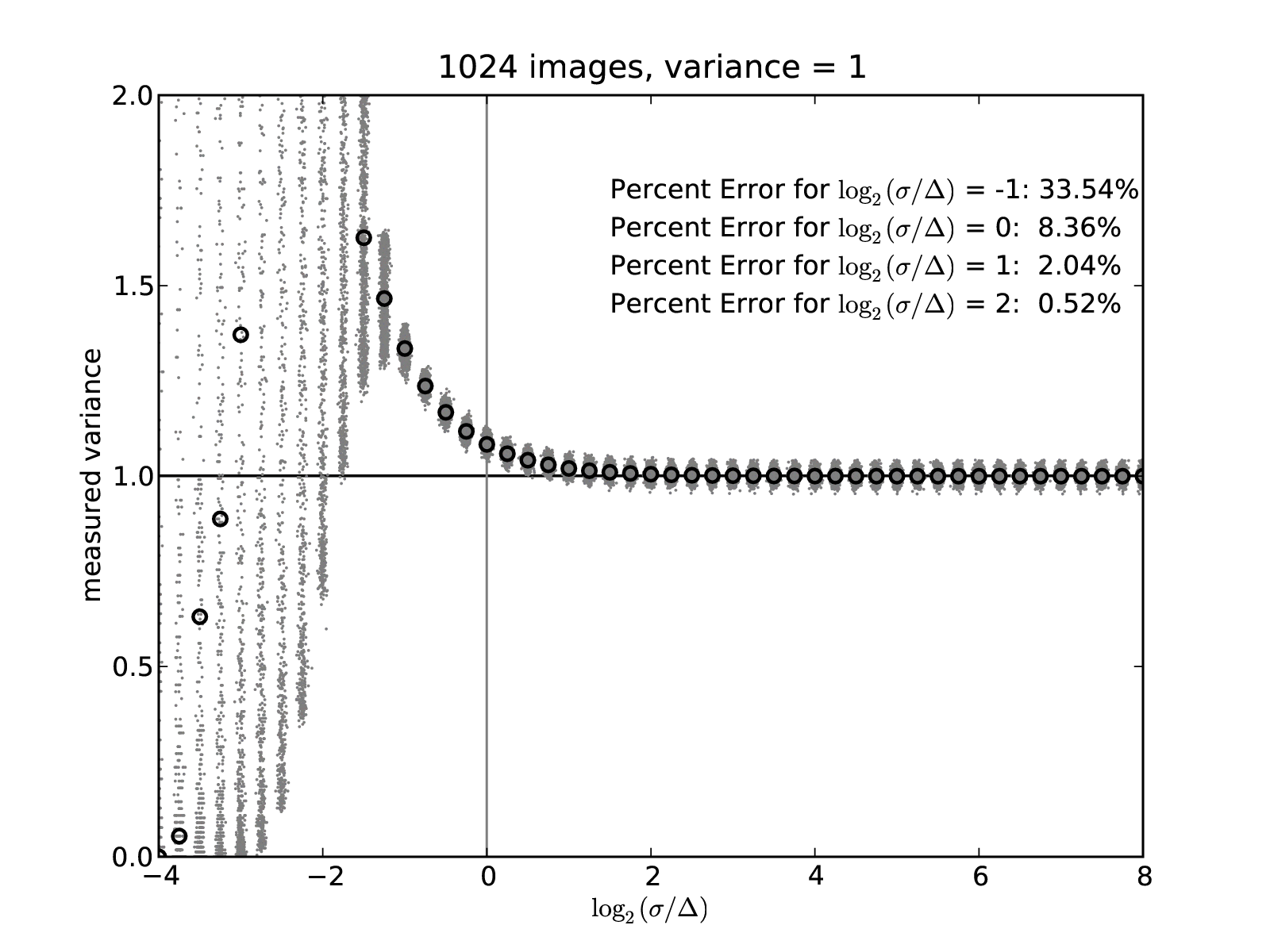}
\caption{Measurement of image noise variance as a function of bit
  depth or minimum representable difference $[\Delta/\sigma]$ for images
  with a randomly chosen mean level and gaussian noise with true
  variance $\sigma^{2} = 1.0$. Each data point has been dithered a
  small amount horizontally to make the distribution visible. Black
  circles show medians (of samples of 1024) for each value of the
  multiplicative factor. The variance is well measured as long as the
  noise root-variance $\sigma$ is twice the minimum representable
  difference $\Delta$. \label{fig:variance}}
\end{figure}

In the first experiments we determine the effect of the minimum
representable difference on the measurement of the variance of the
noise in the image, which we generated as pure Gaussian noise.  For
this experiment we created empty images, added Gaussian noise with
mean $\nu$ (a real value selected in the range 0 to 100) and variance
$\sigma^2 = 1$, and applied the SNIP
procedure. \figurename~\ref{fig:variance} shows the dependance of the
measured variance on the minimum representable difference $[\Delta /
  \sigma]$. As expected, the measured variance increases in accuracy
as the minimum representable difference decreases; the accuracy is
good when $\Delta < 0.5\,\sigma$.

In the second experiments we add a star to each image and ask how well
we can measure its centroid. We added the randomly placed ``star''
before applying the SNIP procedure.  The star is given a set
integrated flux, a FWHM of 2.35~pix, and a randomly selected true
centroid $(x_0,y_0)$ within the central few pixels of the image.  Our
technique for measuring the centroid of the star involves fitting a
quadratic surface to a $3\times 3$ section of the image data with the
center of this array set on a first-guess value for the star
position. It re-centers the $3\times 3$ array around the highest-value
pixel in the neighborhood of the first-guess value. We perform a
simple least-squares fit to these data, using $I(x,y) = a + b x + c y
+ d x^2 + e x y + f y^2$ as our surface model, where $x$ and $y$ are
pixel coordinates in the $3\times 3$ grid, and $a$, $b$, $c$, $d$,
$e$, and $f$ are parameters.  Our centroid measurement $(x_s,y_s)$ is
then computed from the best-fit parameters by
$$
(x_s, y_s) = \left(\frac{c e - b f}{2 d f - 2 e^{2}},
                       \frac{b e - c d}{2 d f - 2 e^{2}}\right)
$$
The offset of this measurement from the true value (in pix) is then
$\sqrt{(x_s - x_0)^{2} + (y_s - y_0)^{2}}$.  For sources strongly
affected by noise, this fitting method sometimes returns large
offsets; we artificially cap all offsets at 2~pix.

In the third experiments we consider the effect of quantization on the
photometric properties of the star. We have now centroided the star,
and so we use the position of the star as found in the above paragraph
along with the known variance $\sigma^2$ to do a Gaussian fit to the
point spread function of the star. We only allow the height $A$ to
vary for fitting the star, which is related to the total flux $S$ of the
star by $S = A \times 2\pi\sigma^2$. The fit is therefore really just
a linear fit which is represented by the model
$Ae^{\frac{-(x-x_s)^2-(y-y_s)^2}{2\sigma^2}} + \mu_{sky}$, where
$\mu_{sky}$ is the ``sky level.''

\begin{figure}
\includegraphics[width=\textwidth]{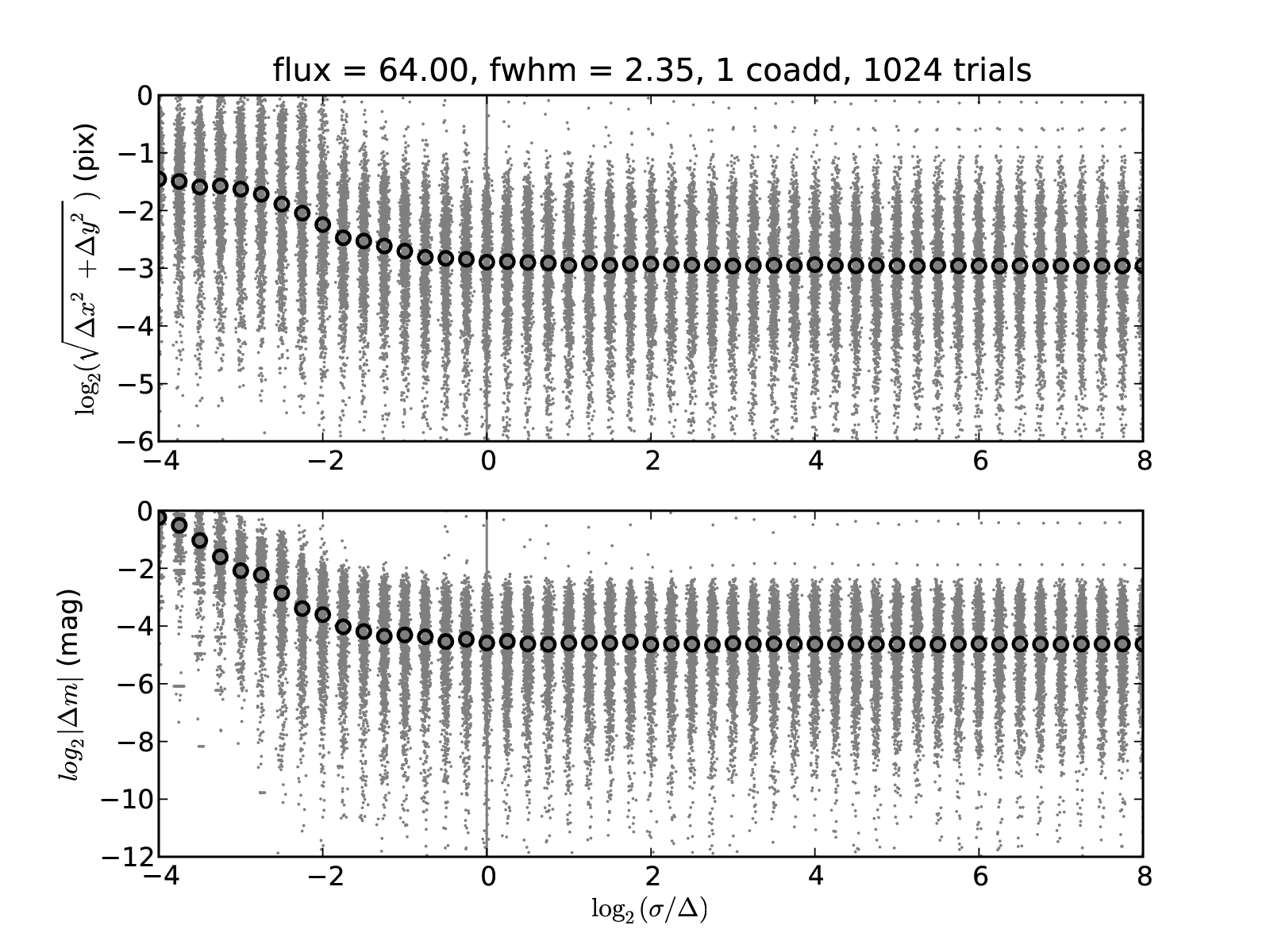}
\caption{For both plots, the black circles show the median values. The
  points are generated by generating 1024 images with noise variance
  $\sigma^{2} = 1.0$ and a gaussian star randomly placed with a total
  flux of 64.0 and a FWHM of 2.35~pix.  The peak per-pixel intensity
  of the star is $10\,\sigma$. \textsl{Top:} Plot of measured star
  offset (astrometric error in pixels; see text for centroiding
  procedure and offset calculation) as a function of bit depth or
  minimum representable difference $[\Delta/\sigma]$.
  \textsl{Bottom:} Plot of $log_2$ of the absolute value of the
  difference between measured magnitude and the true magnitude of the
  star (photometric error; logarithm of the logarithm!)\ as a function
  of bit depth or minimum representable difference
  $[\Delta/\sigma]$.\label{fig:bitsoffset64}}
\end{figure}

\begin{figure}
\includegraphics[width=\textwidth]{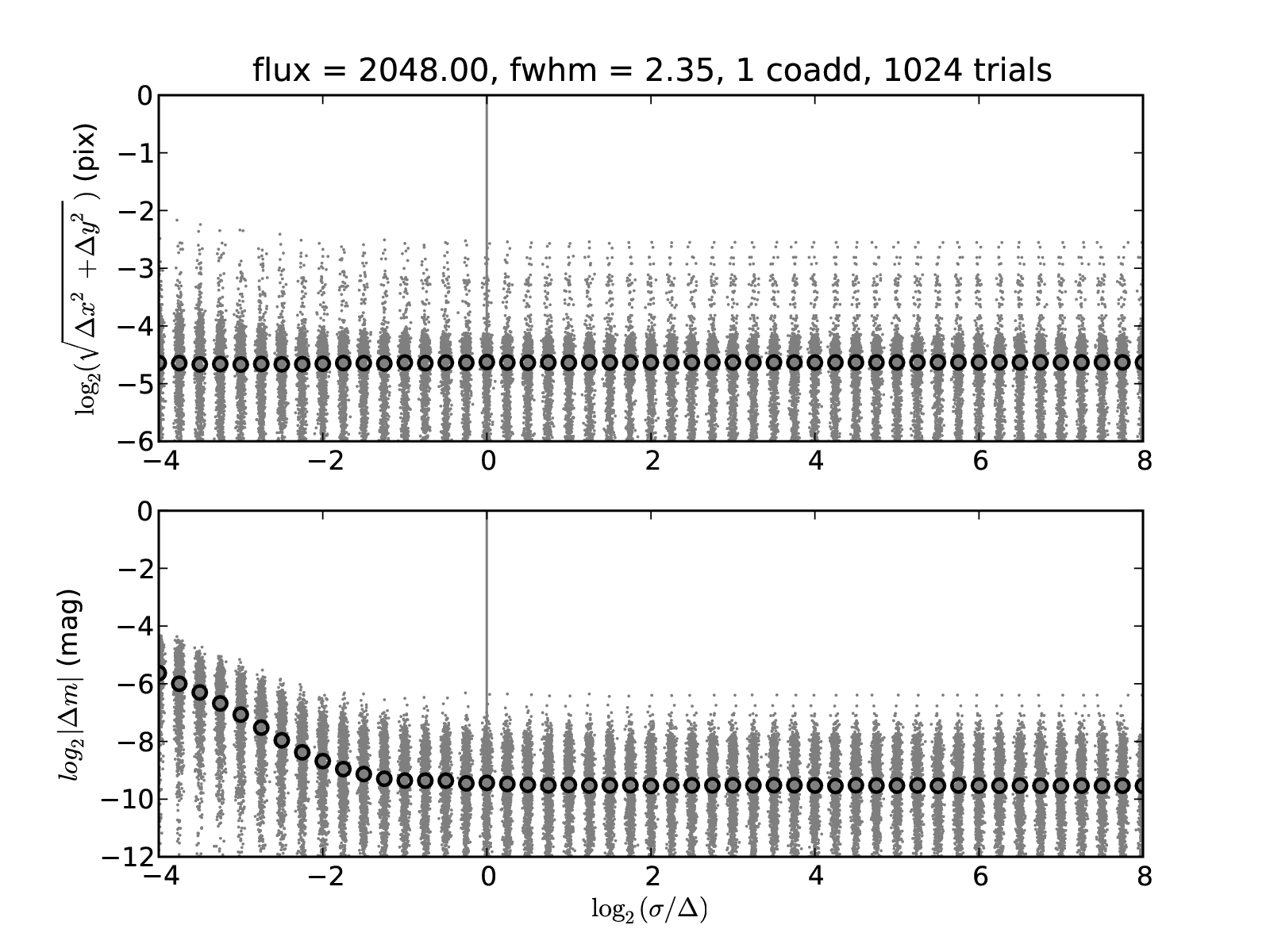}
\caption{Same as \figurename~\ref{fig:bitsoffset64} except the total
  flux of the star was set to 2048.0.  The peak per-pixel intensity of
  the star is now $320\,\sigma$.\label{fig:bitsoffset2048}}
\end{figure}

\figurename~\ref{fig:bitsoffset64} shows the dependence of the
measured centroids and brightnesses of the stars relative to the known
values as a function of minimum representable difference $\Delta$. We
find (not surprisingly) that the accuracy by which we measure the
centroid and brightness increases as $\Delta$ decreases; the accuracy
saturates at $\Delta < 0.5\,\sigma$.  The expectation is that a star
of high flux compared to the noise level will be very accurately
measured, even at the highest minimum representable difference $\Delta
= 16\,\sigma$. At lower fluxes the offset is expected to be
larger. \figurename s~\ref{fig:bitsoffset64} and
\ref{fig:bitsoffset2048} confirm these intuitions.  In each of these
\figurename s, the experiment is performed on 1024 independent
trials---each trial image has a unique sky level, noise sample, and
star position---and each trial image has been SNIPped at each value of
$\Delta$.

\begin{figure}
\includegraphics[width=\textwidth]{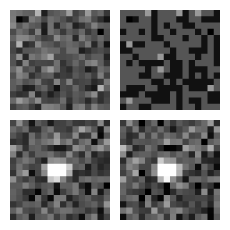}
\caption{Four $16\times 16$ pixel images that demonstrate the coadding
  procedure. The top left image shows a single image with noise
  variance $\sigma^{2}$ = 1.0 and an (extremely faint) Gaussian star
  with a total flux of 2.0 and FWHM of 2.35~pix.  The top right image
  is the same as the top left, but with the pixel values snapped to
  finite resolution $[\sigma/\Delta]=2$ or minimum representable
  difference $\Delta = 0.5\,\sigma$. The bottom left image shows the
  result of coadding 1024 images without snapping to finite resolution.  The
  bottom right image is the same but coadding \emph{after} snapping
  each individual image data to $[\sigma/\Delta]=2$. The similarities of
  the images indicates that information has been preserved.  The peak
  per-pixel intensity of the star is $0.32\,\sigma$; this star is not
  visible in any of the individual images, but appears in the coadded
  images.\label{fig:provecoadd}}
\end{figure}

Tiny variations in mean pixel values, even those smaller than the
noise amplitude, do contain valuable information.  In the fourth
experiments we investigate this by coadding noise-dominated images of
the same region of the sky to reveal sources too faint to be detected
in any individual image.  The test we perform is---for each trial---to take 1024 images,
add a faint source (fainter than any detection limit) to each image at
a common location $(x_0,y_0)$, generate independent sky level $\nu$
and sky noise for each image, apply the SNIP method for the same range
of minimum representable differences $\Delta$, coadd the images and
measure the star offsets in the SNIPped, coadded images.  The star
position remains constant, but each image has independent sky
properties. Just to reiterate, we coadd \emph{after} applying the SNIP
procedure, but given enough images we can measure astrometric and
photometric properties of the extremely faint star with good accuracy.
The coadding procedure is illustrated in \figurename~\ref{fig:provecoadd}.

\figurename~\ref{fig:bitsoffsetcoadd1} shows that for a total star flux of
$2.0$, if we coadd 1024 images with independent sky properties, we can centroid and photometer the source with similar quality to the ``single exposure''
\figurename~\ref{fig:bitsoffset64} with a total star flux of 64.0.  This is expected when the minimum representable difference $\Delta$ is small.
What may not be expected is that even sources for which
every pixel is fainter than the minimum representable difference
$\Delta$ in any individual image pixel, we are able to detect,
centroid, and photometer accurately by coadding images together; that
is, the imposition of a large individual-image minimum representable
difference does not distort information about exceedingly faint
sources.
\figurename~\ref{fig:bitsoffsetcoadd2} shows the same for total flux 64.0; the trend is similar to that in \figurename~\ref{fig:bitsoffset2048}.

In the coadd experiments, we have made the optimistic assumption that the sky level will be independent in every image that contributes to each coadd trial.  To test the
importance of varying the sky among the coadded exposures, we made a
version in which we did \emph{not} vary the sky.  That is, we made each
individual image not just with a fixed star flux and location but also
a fixed sky level---different for each trial, but the same for each coadded exposure within each trial.  The differences between \figurename s~\ref{fig:bitsoffsetcoadd1} and
\ref{fig:bitsoffsetcoadd1_samesky} are substantial when the minimum representable difference $\Delta$ is significantly larger than the per-pixel noise level $\sigma$.

\section{Discussion}

Because of finite noise, the information content in astronomical
images is finite, and can be captured by a finite numerical
resolution.  In the above, we scaled and snapped-to-integer
real-valued images by a SNIP procedure such that in the SNIPped image,
the minimum representable difference $\Delta$ between pixel values was
set to a definite fraction of the Gaussian noise root-variance (sigma)
$\sigma$.  We found with direct numerical experiments that the SNIP
procedure introduces essentially no significant error in estimating
the variance of the image, or in centroiding or photometering stars in
the image, when the minimum representable difference is set to any
value $\Delta \leq 0.5\,\sigma$.  In addition, we showed that all the
information about sources fainter than the per-pixel noise level is
preserved by the quantization (SNIP) procedure, again provided that
$\Delta \leq 0.5\,\sigma$.  This is somewhat remarkable because at
$\Delta = 0.5\,\sigma$ the faintest sources in our experiments were
fainter than the minimum representable difference.

\begin{figure}
\includegraphics[width=\textwidth]{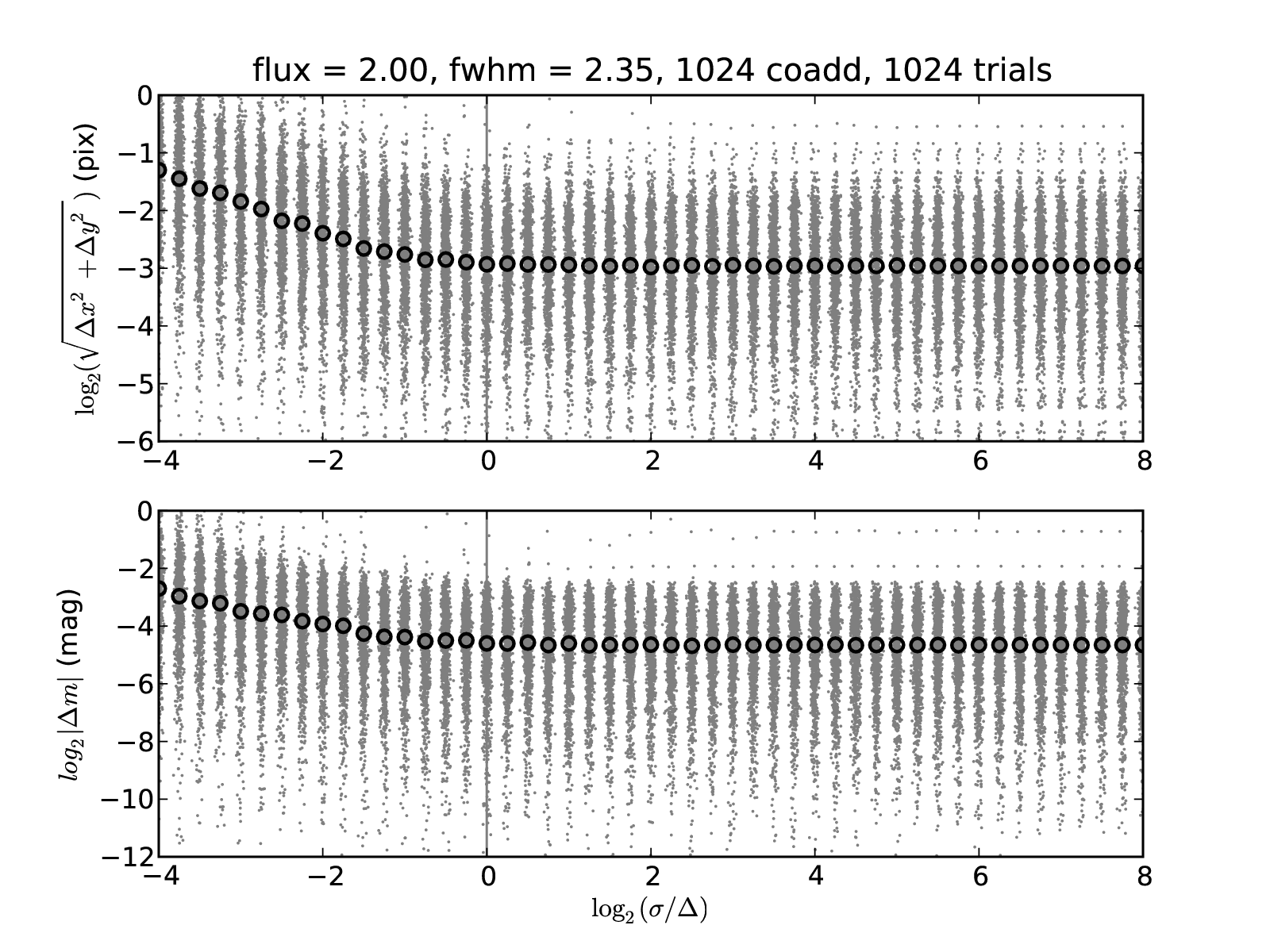}
\caption{Same as \figurename~\ref{fig:bitsoffset64} except with a star
  of total flux 2.0, and coadding sets of 1024 exposures after
  snap-to-integer to make the extremely faint source detectable.  In
  this experiment we give each of the coadded images a different sky
  level (see text). \label{fig:bitsoffsetcoadd1}}
\end{figure}

\begin{figure}
\includegraphics[width=\textwidth]{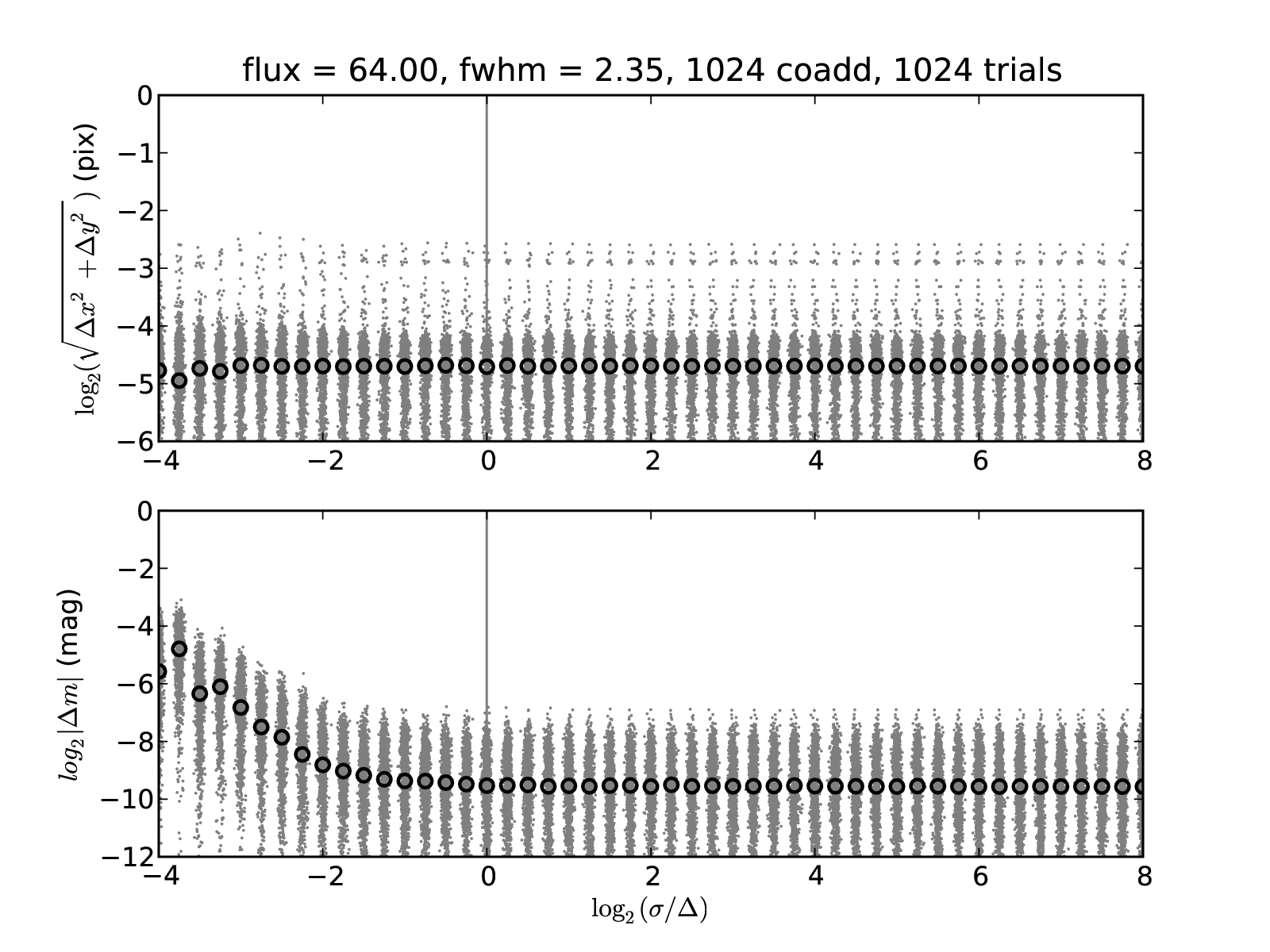}
\caption{Same as \figurename~\ref{fig:bitsoffsetcoadd1} except with a
  star of total flux 64.0. \label{fig:bitsoffsetcoadd2}}
\end{figure}

\begin{figure}
\includegraphics[width=\textwidth]{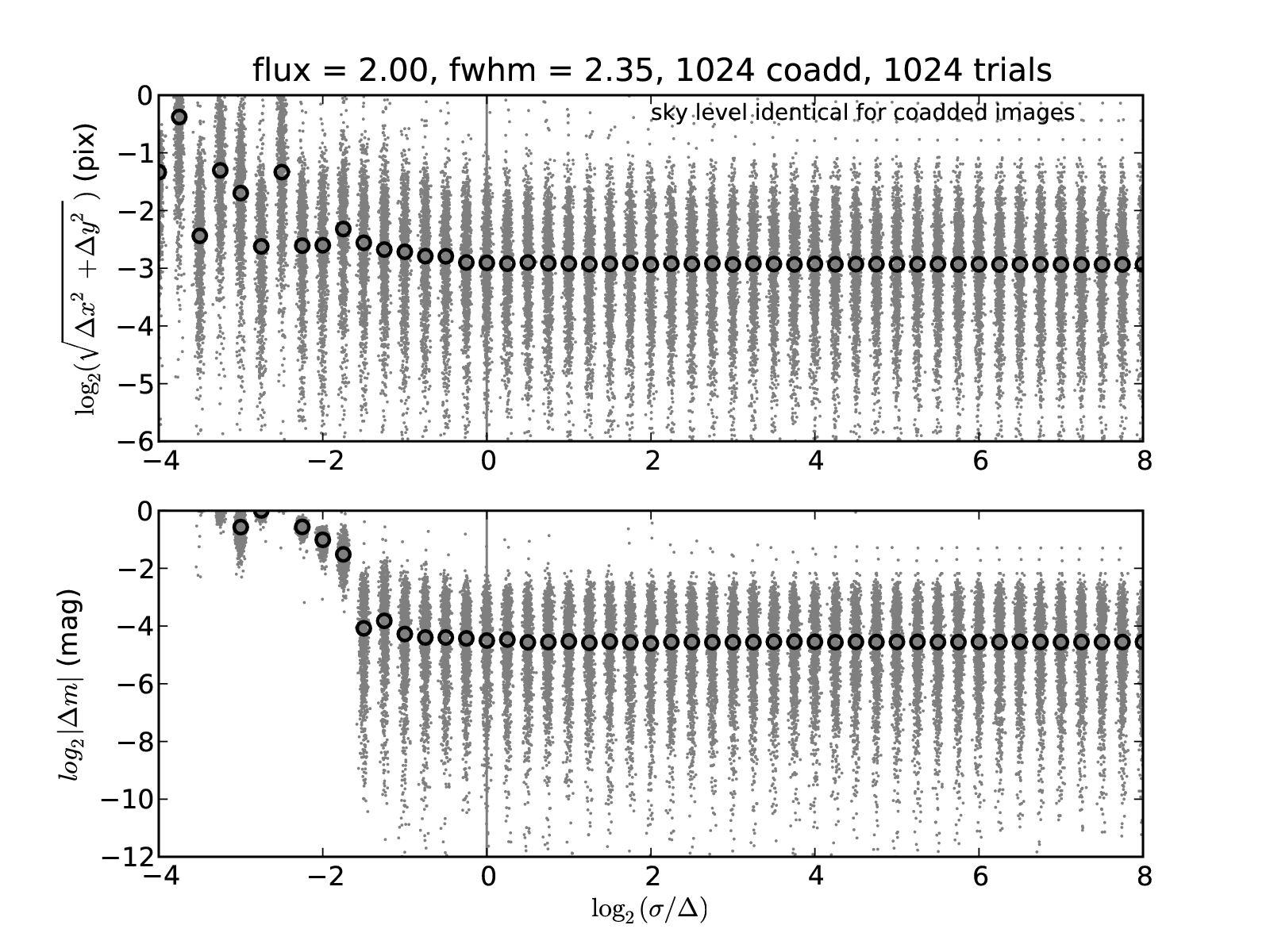}
\caption{Same as \figurename~\ref{fig:bitsoffsetcoadd1} except that in
  this experiment we give each of the coadded images the same sky
  level; the sky level is different for each of the coadd trials, but
  the same for all the images within each coadd
  trial. \label{fig:bitsoffsetcoadd1_samesky}}
\end{figure}

Although it is somewhat counterintuitive that integer quantization of
the data does not remove information about sources fainter than the
quantization level, it is perhaps even more counterintuitive how well
photometric measurements perform in our coadd tests.  For example, in
\figurename s~\ref{fig:bitsoffsetcoadd1} and
\ref{fig:bitsoffsetcoadd2}, the photometric measurements are
relatively accurate even when the data are quantized at minimum
representable difference $\Delta = 16\,\sigma$!  The quality of the
measurements can be understood in part by noting that the coadded
images have a per-pixel noise $\sigma = \sqrt{1024}\,\sigma =
32\,\sigma$, which is once again larger than the minimum representable
difference, and in part by noting that each image has a different sky
level, so each individual image is differently ``wrong'' in its
photometry; many of these differences average out in the coadd.  When
the sky level is held fixed across coadded images, photometric
measurements become inaccurate again---as seen in
\figurename~\ref{fig:bitsoffsetcoadd1_samesky}---because
individual-image biases caused by the coarse quantization no longer
``average out''.

Our fundamental conclusion is that all of the scientifically relevant
information in an astronomical image is preserved as long as the
minimum representable difference $\Delta$ in pixel values is smaller
than or equal to half the per-pixel root-variance (sigma) $\sigma$ in
the image noise.  This confirms previous results based on
information-theory arguments (for example, \citealt{Gaztanaga}), and
extends previous experiments on bright-source photometry
\citep{WhiteGreenfield, PenceInPress} to
astrometry and to sources fainter than the noise.

Our experiments were performed on images with pure Gaussian noise; of
course many images contain significant non-Gaussianity in their
per-pixel noise so the empirical variance will depart significantly
from the true noise variance \citep{WhiteGreenfield}.  The
conservative approach for such images is to take not the true variance
for $\sigma^2$ but rather use for $\sigma^2$ something like the
minimum of the straightforwardly measured variance and a central
variance estimate, such as a sigma-clipped variance estimate, an
estimate based on the curvature of the central part of the noise value
frequency distribution function, or the median absolute difference of
nearby pixels \citep{Pence}.  With this re-definition of the root
variance $\sigma$, the condition $\Delta \leq 0.5\,\sigma$ represents
a conservative setting of the minimum representable difference.

The fact that a $\Delta = 0.5\,\sigma$ representation preserves
information on the faint sources---even those fainter than $\Delta$
itself---has implications for the design of data-taking systems, which
are necessarily limited in bandwidth.  If the system is set with
$\Delta$ substantially smaller than $0.5\,\sigma$, then bright sources
will saturate the representation more frequently than necessary, while
no additional information is being carried about the faintest sources.
Any increase in $\Delta$ pays off directly in putting more of the
necessarily limited system dynamic range onto bright sources, so it
behooves system designers to push as close to the $\Delta =
0.5\,\sigma$ limit as possible.

To put this in the context of a real data system, we looked at a
``DARK'' calibration image from the Hubble Space Telescope Advanced
Camera for Surveys (ACS).  The dark image should have the lowest
per-pixel noise of any ACS image, because it has only dark and read
noise.  We chose image set \texttt{jbanbea2q}, and measured the median
noise level in the raw DARK image with the median absolute difference
between values of nearby pixels (for robustness).  The ACS data system
is operating with a minimum representable difference $0.25\,\sigma <
\Delta < 0.33\,\sigma$, comfortably within the information-preserving
range and close to the minimum-bandwidth limit of $\Delta =
0.5\,\sigma$.  Of course this is for a dark frame; sky exposures
(especially long ones) could have been profitably taken with a larger
$\Delta$ (because $\sigma$ will be greater); this would have preserved
more of the system dynamic range for bright sources.  If the ACS took
almost exclusively long exposures, the output would contain more
scientific information with a larger setting of the minimum
representable difference.

In some sense, the results of this paper recommend a ``lossy'' image
compression technique, in which data are scaled by a factor and
snapped to integer values such that the minimum representable
difference $\Delta$ is made equal to or smaller than $0.5\,\sigma$.
Indeed, when typical real-valued astronomical images are converted to
integers at this resolution, the integer versions compress far better
with subsequent standard file compression techniques (such as gzip)
than do the floating-point originals \citep{Gaztanaga, Watson,
  WhiteGreenfield, Pence, bernstein}.  In the $\Delta = 0.5\,\sigma$
representation, after lossless compression, storage and transmission
of the image ``costs'' only a few bits per noise-dominated pixel.
Because the snap-to-integer step changes the data, this overall
procedure is technically lossy, but we have shown here that none of
the \emph{scientific} information in the image has been lost.

\acknowledgements We thank Mike Blanton, Doug Finkbeiner, and Dustin
Lang for useful discussions, and the anonymous referee for
constructive suggestions.  Support for this work was provided by NASA
(NNX08AJ48G), the NSF (AST-0908357), and the Alexander von~Humboldt
Foundation.

\bibliographystyle{apj}
\bibliography{apj-jour,refs}

\clearpage

\end{document}